\documentclass[preprint]{elsart}

\usepackage{bm}
\usepackage{graphicx, epsfig}
\usepackage{amsmath,amssymb}

\newcommand{\Es}{\mathbb{E}}
\newcommand{\Cov}{{\rm Cov}}
\newcommand{\Var}{{\rm Var}}

\journal{CSDA}

\begin{document}

\begin{frontmatter}
\title{Birnbaum--Saunders nonlinear regression models}
\author{Artur J.~Lemonte}
\address{Departamento de Estat\'istica, Universidade de S\~ao Paulo, Rua do Mat\~ao, 1010,
S\~ao Paulo/SP, 05508-090, Brazil}
\author{Gauss M.~Cordeiro}
\address{Departamento de Estat\'istica e Inform\'atica, Universidade Federal Rural de Pernambuco,
Recife/PE, 52171-900, Brazil}

\begin{abstract}

We introduce, for the first time, a new class of Birnbaum--Saunders nonlinear
regression models potentially useful in lifetime data analysis. The class generalizes the
regression model described by Rieck and Nedelman [1991, A log-linear model for the
Birnbaum--Saunders distribution, {\em Technometrics}, {\bf 33}, 51--60].
We discuss maximum likelihood estimation for the parameters
of the model, and derive closed-form expressions for the
second-order biases of these estimates. Our formulae are easily
computed as ordinary linear regressions and are then used to define bias
corrected maximum likelihood estimates. Some simulation results
show that the bias correction scheme yields nearly unbiased estimates without
increasing the mean squared errors. We also give an application to a real fatigue
data set.

\end{abstract}
\end{frontmatter}

%
\section{Introduction}\label{introduction}

Different regression models have been proposed for lifetime data such as
those based on the gamma, lognormal and Weibull distributions. These models
typi\-cally provide a satisfactory fit in the middle portion of the data, but
very often fail to deliver a good fit at the tails, where only
a few observations are gene\-rally available. The family of distributions proposed
by Birnbaum and Saunders (1969) can also be used to model
lifetime data and it is widely applicable to model failure times
of fatiguing materials. This family has the appealing feature of providing
satisfactory tail fitting. This family of distributions was originally obtained from
a model for which failure follows from the development
and growth of a dominant crack. It was later derived by Desmond (1985) using a
biological model which followed from relaxing some of the assumptions
originally made by Birnbaum and Saunders (1969).

The random variable $T$ is said to be Birnbaum--Saunders distributed with
parameters $\alpha, \eta > 0$, say $\mathcal{B}$-$\mathcal{S}(\alpha, \eta)$,
if its cumulative distribution function (cdf) is given by
\[
F_{T}(t)=\Phi\Biggl[\frac{1}{\alpha}\Biggl(\sqrt{\frac{t}{\eta}}
        -\sqrt{\frac{\eta}{t}}\Biggr)\Biggr],\quad t > 0,
\]
where $\Phi(\cdot)$ is the standard normal distribution function
and $\alpha$ and $\eta$ are shape and scale parameters,
respectively. It is easy to show that $\eta$ is the median of
the distribution: $F_{T}(\eta)=\Phi(0)=1/2$. For any $k>0$, then
$kT \sim\mathcal{B}$-$\mathcal{S}(\alpha, k\eta)$.

McCarter~(1999) considered parameter estimation under type II data
censo\-ring for the $\mathcal{B}$-$\mathcal{S}(\alpha,\eta)$ distribution.
Lemonte et al.~(2007) derived the second-order biases of the maximum
likelihood estimates (MLEs) of $\alpha$ and $\eta$, and obtained a
corrected likelihood ratio statistic for testing the parameter
$\alpha$. Lemonte et al.~(2008) proposed several bootstrap bias corrected
estimates of $\alpha$ and $\eta$. Further details on the Birnbaum--Saunders
distribution can be found in Johnson et al.~(1995).

Rieck and Nedelman (1991) proposed a log-linear regression model based on the
Birnbaum--Saunders distribution. They showed that if
$T\sim\mathcal{B}$-$\mathcal{S}(\alpha,\eta)$, then $Y=\log(T)$ is
sinh-normal distributed, say $Y\sim\mathcal{SN}(\alpha,\mu,\sigma)$,
with shape, location and scale parameters given by $\alpha$, $\mu=\log(\eta)$ and
$\sigma=2$, respectively. Their model has been widely used as an alternative model to
the gamma, lognormal and Weibull regression models; see Rieck and
Nedelman (1991, \S~7). Diagnostic tools for the Birnbaum--Saunders
regression model were developed by Galea et al.~(2004),
Leiva et al.~(2007) and Xie and Wei (2007), and the Bayesian inference was
considered by Tisionas (2001).

In this paper we propose a class of Birnbaum--Saunders nonlinear
regression models which generalizes the regression model introduced by
Rieck and Nedelman (1991). We discuss maximum likelihood estimation of
the regression parameters and obtain the Fisher information matrix. As
is well known, however, the MLEs, although consistent, are typically
biased in finite samples. In order to overcome this shortcoming, we derive a
closed-form expression for the bias of the MLE in these models which is
used to define a bias corrected estimate.

Bias adjustment has been extensively studied in the statistical literature.
In fact, Cook et al. (1986) proposed bias correction in normal nonlinear models.
Young and Bakir (1987) obtained bias corrected estimates for a generalized
log-gamma regression model. Cordeiro and McCullagh (1991)
gave general matrix formulae for bias correction in generalized linear models,
whereas Paula (1992) derive the second-order biases in exponential
family nonlinear mo\-dels. Cordeiro et al.~(2000)
obtained bias correction for symmetric nonlinear regression models.
More recently, Vasconcellos and Cribari--Neto (2005) calculate the biases
of the MLEs in a new class of beta regression. Cordeiro and Dem\'etrio (2008)
propose formulae for the second-order biases of the maximum quasi-likelihood
estimates, whereas Cordeiro and Toyama (2008) derive the second-order biases
in ge\-neralized nonlinear models with dispersion covariates.

The rest of the paper is as follows. Section~\ref{reg_nonlinear}
introduces the class of Birnbaum-Saunders nonlinear regression
models and discusses maximum likelihood estimation. Using general results
from Cox and Snell (1968), we derive in Section~\ref{BIAS}
the second-order biases of the MLEs of the nonlinear parameters in
our class of models and define bias corrected estimates.
Some special models are consi\-de\-red in Section 4. Simulation
results are presented and discussed in Section~\ref{simulation}
for two nonlinear regression models. We show that the bias corrected estimates
are nearly unbiased with mean squared errors very close to the
corresponding ones of the uncorrected estimates. Section~\ref{application}
gives an application of the proposed regression model to a real fatigue data set,
which provides a better fit at the tail of the data. Finally,
Section~\ref{conclusions} concludes the paper.

%

\section{Model specification}\label{reg_nonlinear}

Let $T\sim\mathcal{B}$-$\mathcal{S}(\alpha, \eta)$. The
density function of $Y=\log(T)\sim\mathcal{SN}(\alpha,\mu,\sigma)$
has the form (Rieck and Nedelman, 1991)
\[
\pi(y;\alpha,\mu,\sigma)=\frac{2}{\alpha\sigma\sqrt{2\pi}}\cosh\biggl(\frac{y -
\mu}{\sigma}\biggr)\exp\biggl\{-\frac{2}{\sigma^2}
\mathrm{sinh}^2\biggl(\frac{y-\mu}{\sigma}\biggr)\biggr\}, \quad y\in\mathrm{I\!R}.
\]
This distribution has a number of interesting properties (Rieck, 1989):
(i) It is symmetric around the location parameter $\mu$; (ii) It is unimodal for
$\alpha\leq 2$ and bimodal for $\alpha > 2$; (iii) The mean and
variance of $Y$ are $\Es(Y)=\mu$ and Var$(Y)=\sigma^2 w(\alpha)$,
respectively. There is no closed-form expression for $w(\alpha)$,
but Rieck (1989) obtained asymptotic approximations for both small
and large values of $\alpha$; (iv) If $Y_{\alpha}\sim\mathcal{SN}(\alpha,\mu,\sigma)$,
then $S_{\alpha}=2(Y_{\alpha}-\mu)/(\alpha\sigma)$ converges in distribution
to the standard normal distribution when $\alpha\to 0$.

We define the nonlinear regression model
\begin{equation}\label{eq1}
y_{i}= f_{i}(\bm{x}_{i};\bm{\beta})+\varepsilon_{i},\quad i = 1,\ldots,n,
\end{equation}
where $y_{i}$ is the logarithm of the $i$th observed lifetime,
$\bm{x}_{i}$ is an $m\times 1$ vector of known explanatory variables
associated with the $i$th observable response $y_{i}$,
$\bm{\beta}=(\beta_1,\ldots,\beta_p)^{\top}$ is a vector of
unknown nonlinear parameters, and $\varepsilon_{i}\sim\mathcal{SN}(\alpha,0,2)$.
We assume a nonlinear structure for the location
parameter $\mu_{i}$ in model~(\ref{eq1}), say $\mu_{i}=f_{i}(\bm{x}_{i};\bm{\beta})$, where
$f_{i}$ is assumed to be a known and twice continuously differentiable
function with respect to $\bm{\beta}$. For the linear regression
$\mu_{i}=\bm{x}_{i}^{\top}\bm{\beta}$, the model~(\ref{eq1})
reduces to Rieck and Nedelman's (1991) model.

The log-likelihood function for the vector parameter $\bm{\theta}=(\bm{\beta}^{\top},\alpha)^{\top}$
from a random sample $\bm{y}=(y_1,\ldots,y_n)^{\top}$ obtained from~(\ref{eq1}), except for
constants, can be expressed as
\begin{equation}\label{eq2}
\ell(\bm{\theta})=\sum_{i=1}^{n}\log(\xi_{i1}) - \frac{1}{2}\sum_{i=1}^{n}\xi_{i2}^{2},
\end{equation}
where $\xi_{i1}=\xi_{i1}(\bm{\theta})= 2\alpha^{-1}\cosh([y_i-\mu_{i}]/2)$,
$\xi_{i2}=\xi_{i2}(\bm{\theta})=2\alpha^{-1}\sinh([y_i-\mu_{i}]/2)$
for $i=1,\ldots,n$. The function $\ell(\bm{\theta})$ is assumed to be
regular (Cox and Hinkley, 1974, Ch.~9) with respect to all $\bm{\beta}$
and $\alpha$ derivatives up to third order. Further, the $n\times p$
local matrix
$\bm{D}=\bm{D}(\bm{\beta})=\partial\bm{\mu}/\partial\bm{\beta}$
of partial derivatives of $\bm{\mu}=(\mu_{1},\ldots,\mu_{n})^{\top}$ with
respect to $\bm{\beta}$ is assumed to be of full rank,
i.e., rank($\bm{D})=p$ for all $\bm{\beta}$. The
nonlinear predictors $\bm{x}_1,\ldots,\bm{x}_n$ are embedded in an
infinite sequence of $m \times 1$ vectors that must satisfy these
re\-gu\-larity conditions for the asymptotics to be valid. Under
these assumptions, the MLEs have good asymptotic properties such as
consistency, sufficiency and normality.

The derivatives with respect to the components of $\bm{\beta}$
and $\alpha$ are denoted by: $U_{r} = \partial\ell(\bm{\theta})/\partial\beta_{r}$,
$U_{\alpha}=\partial\ell(\bm{\theta})/\partial\alpha$,
$U_{rs}=\partial^{2}\ell(\bm{\theta})/\partial\beta_{r}\partial\beta_{s}$,
$U_{r\alpha}=\partial^{2}\ell(\bm{\theta})/\partial\beta_{r}\partial\alpha$,
$U_{rst}=\partial^{3}\ell(\bm{\theta})/\partial\beta_{r}\partial\beta_{s}\partial\beta_{t}$,
$U_{rs\alpha}=\partial^{3}\ell(\bm{\theta})/\partial\beta_{r}\partial\beta_{s}\partial\alpha$, etc.
Further, we use the following notation for joint cumulants of log-likelihood
derivatives: $\kappa_{rs} = \Es(U_{rs})$, $\kappa_{r,\alpha} = \Es(U_{r}U_{\alpha})$,
$\kappa_{rst} = \Es(U_{rst})$, etc. Let $\kappa_{rs}^{(t)}=\partial\kappa_{rs}/\partial\beta_{t}$, etc.
All $\kappa$'s and their derivatives are assumed to be of order $\mathcal{O}(n)$.
Also, we adopt the notation $d_{ir} = \partial \mu_{i}/\partial\beta_{r}$
and $g_{irs}=\partial^2 \mu_{i}/\partial\beta_{r}\partial\beta_{s}$ for the first
and second partial derivatives of $\mu_{i}$ with respect to the elements of $\bm{\beta}$.

It is easy to see by differentiating~(\ref{eq2}) that
\[
U_{r}=\frac{1}{2}\sum_{i=1}^{n}d_{ir}\biggl(\xi_{i1}\xi_{i2} - \frac{\xi_{i2}}{\xi_{i1}}\biggr),\quad
U_{\alpha}=-\frac{n}{\alpha} + \frac{1}{\alpha}\sum_{i=1}^{n}\xi_{i2}^{2},
\]
\[
U_{rs} = \frac{1}{2}\sum_{i=1}^{n}g_{irs}\biggl(\xi_{i1}\xi_{i2} - \frac{\xi_{i2}}{\xi_{i1}}\biggr)
-\frac{1}{4}\sum_{i=1}^{n}d_{ir}d_{is}\biggl(2\xi_{i2}^2+\frac{4}{\alpha^2}-1+ \frac{\xi_{i2}^2}{\xi_{i1}^2}\biggr),
\]
\[
U_{r\alpha} = -\frac{1}{\alpha}\sum_{i=1}^{n}d_{ir}\xi_{i1}\xi_{i2}\quad{\rm and}\quad
U_{\alpha\alpha}=\frac{n}{\alpha^2} -\frac{3}{\alpha^2}\sum_{i=1}^{n}\xi_{i2}^{2}.
\]
The score function for $\bm{\beta}$ is
$\bm{U}_{\bm{\beta}}=\frac{1}{2}\bm{D}^{\top}\!\bm{s}$,
where $\bm{s}=\bm{s}(\bm{\theta})$ is an $n$-vector whose $i$th
element is equal to $\xi_{i1}\xi_{i2}-\xi_{i2}/\xi_{i1}$.

It is well-known that, under general regularity conditions (Cox and Hin\-kley,
1974, Ch.~9), the MLEs are consistent, asymptotically efficient and
asymptotically normal. Let $\widehat{\bm{\theta}}=(\widehat{\bm{\beta}}{\vspace{-1cm}}^{\top},
\widehat{\alpha})^{\top}$ be the MLE of $\bm{\theta}=(\bm{\beta}^{\top},\alpha)^{\top}$.
We can write
$\widehat{\bm{\theta}}\stackrel{a}{\sim}\mathcal{N}_{p+1}(\bm{\theta},
\bm{K}_{\bm{\theta}}^{-1})$ for $n$ large, where $\stackrel{a}{\sim}$
denotes approximately distributed, $\bm{K}_{\bm{\theta}}$ is the block-diagonal
Fisher information matrix given by
$\bm{K}_{\bm{\theta}}=\mathrm{diag}\{\bm{K}_{\bm{\beta}},\kappa_{\alpha, \alpha}\}$,
$\bm{K}_{\bm{\theta}}^{-1}$ is its inverse,
$\bm{K}_{\bm{\beta}}=\psi_{1}(\alpha)(\bm{D}^{\top}\bm{D})/4$ is the
information matrix for $\bm{\beta}$ and $\kappa_{\alpha,\alpha}=2n/\alpha^2$ is the
information for $\alpha$. Also,
\[
\psi_{1}(\alpha)=2+\frac{4}{\alpha^{2}} - \frac{\sqrt{2\pi}}{\alpha}
\biggl\{1 - \mathtt{erf}\biggl(\frac{\sqrt{2}}{\alpha}\biggr)\biggr\}
            \exp\biggl(\frac{2}{\alpha^2}\biggr),
\]
where ${\tt erf}(\cdot)$ is the error function given by
\[
\mathtt{erf}(x)=\frac{2}{\sqrt{\pi}}\int_{0}^{x}\mathrm{e}^{-t^2}\mathrm{d}t.
\]
Details on $\mathtt{erf}(\cdot)$ can be found in Gradshteyn and Ryzhik (2007).
Since $\bm{K}_{\bm{\theta}}$ is block-diagonal, the vector
$\bm{\beta}$ and the scalar $\alpha$ are globally orthogonal (Cox and Reid, 1987) and
$\widehat{\bm{\beta}}$ and $\widehat{\alpha}$ are asymptotically independent.
It can be shown (Rieck, 1989) that $\psi_1(\alpha)\approx 1+4/\alpha^2$ for
$\alpha$ small and $\psi_{1}(\alpha)\approx 2$ for $\alpha$ large.

The MLE $\widehat{\bm{\theta}}$ satisfies $p+1$ equations $U_{r}=U_{\alpha}=0$ for
the components of $\bm{\beta}$ and $\alpha$. The Fisher scoring method can be used to
estimate $\bm{\beta}$ and $\alpha$ simultaneously by iteratively solving the equations
\begin{align*}
\bm{\beta}^{(m+1)} &= \bm{\beta}^{(m)} + (\bm{D}^{(m)\top}\bm{D}^{(m)})^{-1}\bm{D}^{(m)\top}\bm{\zeta}^{(m)},\\
\alpha^{(m+1)} &= \frac{1}{2}\alpha^{(m)}(1 + \bar{\xi}_{2}^{(m)}),
\end{align*}
where $\bm{\zeta}^{(m)} = 2\bm{s}^{(m)}/\psi_{1}(\alpha^{(m)})$ and
$\bar{\xi}_{2}^{(m)} = \sum_{i=1}^{n}\xi_{i2}^{2(m)}/n$ for $m = 0,1,2,\ldots$.

The above equations show that any software with a weighted linear regression
routine can be used to calculate the MLEs of $\bm{\beta}$ and $\alpha$
iteratively. Initial approximations $\bm{\beta}^{(0)}$
and $\alpha^{(0)}$ for the iterative algorithm are used to evaluate
$\bm{D}^{(0)},\bm{\zeta}^{(0)}$ and $\bar{\xi}_{2}^{(0)}$ from which these
equations can be used to obtain the next estimates $\bm{\beta}^{(1)}$ and
$\alpha^{(1)}$. These new values can update $\bm{D},\bm{\zeta}$ and
$\bar{\xi}_{2}$ and so the iterations continue until convergence is achieved.


\section{Biases of estimates of $\bm{\beta}$ and $\alpha$}\label{BIAS}

We now obtain some joint cumulants of log-likelihood derivatives and
their derivatives:
\[
\kappa_{rs} = -\frac{\psi_{1}(\alpha)}{4}\sum_{i=1}^{n}d_{ir}d_{is},\quad
\kappa_{r\alpha} = \kappa_{r\alpha\alpha} = 0,\quad\kappa_{\alpha\alpha} = -\frac{2n}{\alpha^2},\quad
\kappa_{\alpha\alpha\alpha} = \frac{10n}{\alpha^3},
\]
\[
\kappa_{rst} = -\frac{\psi_{1}(\alpha)}{4}\sum_{i=1}^{n}(g_{irs}d_{it} + g_{irt}d_{is} + g_{ist}d_{ir}),\quad
\kappa_{rs\alpha} = \frac{(2+\alpha^2)}{\alpha^3}\sum_{i=1}^{n}d_{ir}d_{is},
\]
\[
\kappa_{rs}^{(t)}=-\frac{\psi_{1}(\alpha)}{4}\sum_{i=1}^{n}(g_{irt}d_{is}+g_{ist}d_{ir}),\quad
\kappa_{r\alpha}^{(\alpha)}=\kappa_{r\alpha}^{(s)}=0\quad{\rm and}\quad
\kappa_{\alpha\alpha}^{(\alpha)}=\frac{4n}{\alpha^3}.
\]

Let $B(\widehat{\beta}_{a})$ and $B(\widehat{\alpha})$ be
the $n^{-1}$ biases of $\widehat{\beta}_{a}$ ($a=1,\ldots,p$)
and $\widehat{\alpha}$, respectively. The use of Cox and Snell's (1968)
formula to obtain these biases is greatly simplified, since $\bm{\beta}$
and $\alpha$ are globally orthogonal and the cumulants corresponding
to the parameters in $\bm{\beta}$ are invariant under permutation of
these parameters. From now on we use Einstein summation convention
with the indices varying over the corresponding parameters. We have
\begin{equation}\label{vies-beta}
B(\widehat{\beta}_{a}) = \sideset{}{^{\prime}}\sum_{s,t,u}\kappa^{a,s}\kappa^{t,u}\biggl(
\kappa_{st}^{(u)}-\frac{1}{2}\kappa_{stu}\biggr) + \kappa^{\alpha,\alpha}
\sideset{}{^{\prime}}\sum_{s}\kappa^{a,s}\biggl(
\kappa_{s\alpha}^{(\alpha)}-\frac{1}{2}\kappa_{s\alpha\alpha}\biggr)
\end{equation}
and
\begin{equation}\label{vies-alpha}
B(\widehat{\alpha}) = (\kappa^{\alpha,\alpha})^{2}\biggl(
\kappa_{\alpha\alpha}^{(\alpha)}-\frac{1}{2}\kappa_{\alpha\alpha\alpha}\biggr) +
\kappa^{\alpha,\alpha}\sideset{}{^{\prime}}\sum_{t,u}\kappa^{t,u}\biggl(
\kappa_{\alpha t}^{(u)}-\frac{1}{2}\kappa_{\alpha tu}\biggr),
\end{equation}
where $\kappa^{r,s}$ is the $(r,s)$th element of the inverse $\bm{K}_{\bm{\beta}}^{-1}$
of the information matrix for $\bm{\beta}$, $\kappa^{\alpha,\alpha}=\kappa_{\alpha,\alpha}^{-1}$
and $\sum^{'}$ denotes the summation over all combinations of parameters
$\beta_{1},\beta_{2},\ldots,\beta_{p}$.

First, we consider equation~(\ref{vies-beta}) from which we
readily have that the second sum is zero
since $\kappa_{s\alpha\alpha}=\kappa_{s \alpha}^{(\alpha)}=0$.
It follows that
\[
B(\widehat{\beta}_{a}) = -\frac{\psi_{1}(\alpha)}{8}\sideset{}{^{\prime}}\sum_{s,t,u}\kappa^{a,s}\kappa^{t,u}
\sum_{i=1}^{n}(g_{isu}d_{it} - g_{ist}d_{iu} + g_{itu} d_{is}).
\]
By rearranging the summation terms we obtain
\[
B(\widehat{\beta}_{a}) = -\frac{\psi_{1}(\alpha)}{8}\sum_{i=1}^{n}
\sideset{}{^{\prime}}\sum_{s}\kappa^{a,s}d_{is}\sideset{}{^{\prime}}\sum_{t,u}\kappa^{t,u}g_{itu}.
\]
Let $\bm{d}_{i}^{\top}$ $(1\times p)$ and $\bm{g}_{i}^{\top}$ $(1\times p^2)$
be vectors containing the first and second partial derivatives of the mean $\mu_{i}$ with
respect to the $\beta$'s. We can write the above equation in matrix notation as
\[
B(\widehat{\beta}_{a}) = -\frac{\psi_{1}(\alpha)}{8}
\bm{\rho}_{a}^{\top}\bm{K}_{\bm{\beta}}^{-1}\sum_{i=1}^{n}\bigl\{\bm{d}_{i}
\bm{g}_{i}^{\top}\bigr\}{\rm vec}(\bm{K}_{\bm{\beta}}^{-1}),
\]
where $\bm{\rho}_{a}^{\top}$ is the $a$th row of the $p\times p$ identity matrix
and vec$(\cdot)$ is the operator which transforms a matrix into a vector by
stacking the columns of the matrix one underneath the other.
It is straightforward to check that
\[
B(\widehat{\beta}_{a}) = -\frac{\psi_{1}(\alpha)}{8}
\bm{\rho}_{a}^{\top}\bm{K}_{\bm{\beta}}^{-1}\bm{D}^{\top}\bm{G}{\rm vec}(\bm{K}_{\bm{\beta}}^{-1}),
\]
where $\bm{D}=\partial\bm{\mu}/\partial\bm{\beta}=(\bm{d}_{1},\ldots,\bm{d}_{n})^{\top}$
and $\bm{G}=\partial^2\bm{\mu}/\partial\bm{\beta}^{\top}\partial\bm{\beta}=(\bm{g}_{1},\ldots,
\bm{g}_{n})^{\top}$ are $n\times p$ and $n\times p^2$ matrices of the
first and second partial derivatives of the mean vector $\bm{\mu}$ with respect
to $\bm{\beta}$, respectively. The $n^{-1}$ bias vector
$\bm{B}(\widehat{\bm{\beta}})$ of $\widehat{\bm{\beta}}$ can then be written as
\begin{equation}\label{bias-beta}
\bm{B}(\widehat{\bm{\beta}})=(\bm{D}^{\top}\bm{D})^{-1}\bm{D}^{\top}\bm{d},
\end{equation}
where $\bm{d}$ is an $n\times 1$ vector defined as
$\bm{d}=-\{2/\psi_{1}(\alpha)\}\bm{G}{\rm vec}\{(\bm{D}^{\top}\bm{D})^{-1}\}$.

We now calculate the $n^{-1}$ bias of $\widehat{\alpha}$.
Using~(\ref{vies-alpha}), we obtain
\begin{align*}
B(\widehat{\alpha})&=-\frac{\alpha}{4n}-\frac{(2+\alpha^2)}{4\alpha n}
\sum_{i=1}^{n}\sideset{}{^{\prime}}\sum_{t,u}\kappa^{t,u}d_{it}d_{iu}
= -\frac{\alpha}{4n}-\frac{(2+\alpha^2)}{4\alpha n}
\sum_{i=1}^{n}\bm{d}_{i}^{\top}\bm{K}_{\bm{\beta}}^{-1}\bm{d}_{i}\\
&= -\frac{\alpha}{4n}-\frac{(2+\alpha^2)}{4\alpha n}{\rm tr}(\bm{D}\bm{K}_{\bm{\beta}}^{-1}\bm{D}^{\top}),
\end{align*}
where $\mathrm{tr}(\cdot)$ denotes the trace operator. Now, making use of the fact
that ${\rm tr}(\bm{D}\bm{K}_{\bm{\beta}}^{-1}\bm{D}^{\top})=4p/\psi_{1}(\alpha)$,
we can rewrite the $n^{-1}$ bias of $\widehat{\alpha}$ as
\begin{equation}\label{bias-alpha}
B(\widehat{\alpha}) = -\frac{1}{n}\biggl\{p\biggl(\frac{2+\alpha^2}{\alpha\psi_{1}(\alpha)}\biggr)
+\frac{\alpha}{4}\biggr\}.
\end{equation}

Equations~(\ref{bias-beta}) and~(\ref{bias-alpha}) represent the main results
of the paper. The bias vector $\bm{B}(\widehat{\bm{\beta}})$ can be
obtained from a simple ordinary least-squares regression of $\bm{d}$ on the columns
of $\bm{D}$. It depends on the nonlinearity of the regression function $f$
and the parameter $\alpha$. The bias vector $\bm{B}(\widehat{\bm{\beta}})$
will be small when $\bm{d}$ is orthogonal to the columns of $\bm{D}$.
Also, it can be large when $\psi_{1}(\alpha)$ and $n$ are both
small. Equation~(\ref{bias-beta}) is easily handled algebraically for
any type of nonlinear regression, since it involves simple operations on
matrices and vectors. For special models with closed-form information matrix for
$\bm{\beta}$, it is possible to obtain closed-form expressions for
$\bm{B}(\widehat{\bm{\beta}})$. For linear models, the matrix $\bm{G}$
and the vector $\bm{d}$ vanish and hence $\bm{B}(\widehat{\bm{\beta}})=\bm{0}$,
which is in agreement with the result due to Rieck and Nedelman (1991, p.~54)
that the MLEs are unbiased to order $n^{-1}$.
Expression (6) depends directly on the nonlinear structure
of the regression model only through the rank $p$ of $\bm{D}$.
It shows that the bias is always a linear function of the
dimension $p$ of $\bm{\beta}$.

In the right-hand sides of expressions~(\ref{bias-beta}) and~(\ref{bias-alpha}),
which are both of order $n^{-1}$, consistent estimates of the parameters
$\bm{\beta}$ and $\alpha$ can be inserted to define bias corrected estimates
$\widetilde{\bm{\beta}}=\widehat{\bm{\beta}}-\widehat{\bm{B}}(\widehat{\bm{\beta}})$
and $\widetilde{\alpha}=\widehat{\alpha}-\widehat{B}(\widehat{\alpha})$,
where $\widehat{\bm{B}}(\widehat{\bm{\beta}})$ and $\widehat{B}(\widehat{\alpha})$ are the values of
$\bm{B}(\widehat{\bm{\beta}})$ and $B(\widehat{\alpha})$, respectively, at
$\widehat{\bm{\theta}}=(\widehat{\bm{\beta}}{\vspace{-1cm}}^{\top},\widehat{\alpha})^{\top}$.
The bias corrected estimates $\widetilde{\bm{\beta}}$ and $\widetilde{\alpha}$ are
expected to have better sampling properties than the classical MLEs $\widehat{\bm{\beta}}$
and $\widehat{\alpha}$. In fact, we present some simulations in Section~\ref{simulation}
to show that $\widetilde{\bm{\beta}}$ and $\widetilde{\alpha}$ have smaller
biases than their corresponding uncorrected estimates, thus suggesting that
these bias corrections have the effect of shrinking the adjusted estimates
toward to the true parameter values. However, we can not say that the bias
corrected estimates offer always some improvement over the MLEs, since they
can have mean squared errors larger.

It is worth emphasizing that there are other methods to obtain bias corrected
estimates. In regular parametric problems, Firth (1993) developed the
so-called ``preventive'' method, which also allows for the removal of the
second-order bias. His method consists of modifying the original score
function to remove the first-order term from the asymptotic bias of these estimates.
In exponential families with canonical parameterizations, his correction scheme
consists in penalizing the likelihood by the Jeffreys invariant priors.
This is a preventive approach to bias adjustment which has its merits,
but the connections between our results and his work are not pursued in this
paper since they could be developed in future research. Additionally,
it should be mentioned that it is possible to avoid cumbersome and
tedious algebra on cumulant calculations by using Efron's bootstrap (Efron and Tibshirani, 1993).
We use the analytical approach here since this leads to a nice formula. Moreover, the application
of the analytical bias approximation seems to generally be the most feasible procedure
to use and it continues to receive attention in the literature.

We now calculate the second-order bias $B(\widehat\mu_{i})$
of the MLE $\widehat\mu_{i}$ of the $i$th mean $\mu_{i}=f_{i}(\bm{x}_{i};\bm{\beta})$.
We can easily show by Taylor series expansion that
\[
B(\widehat{\mu}_{i})= \bm{d}_{i}^{\top} \bm{B}(\widehat{\bm{\beta}})+\frac{1}{2}\mbox{tr}\{\bm{M}_i
\textrm{Cov}(\widehat{\bm{\beta}})\},
\]
where $\bm{M}_i$ is a $p\times p$ matrix of second partial derivatives $\partial^{2}\mu_{i}/
\partial\beta_{r}\partial\beta_{s}$ (for
$r,s=1,\ldots,p$), $\textrm{Cov}(\widehat{\bm{\beta}})=\bm{K}_{\bm{\beta}}^{-1}$
is the asymptotic covariance matrix of $\widehat{\bm{\beta}}$ and the vectors
$\bm{d}_{i}$ and $\bm{B}(\widehat{\bm{\beta}})$ were mentioned previously.
All quantities in the above equation should be evaluated at $\widehat{\bm{\beta}}$.

The asymptotic variance of $\widehat{\mu}_{i}$ can also be expressed explicitly
in terms of the covariance of $\widehat{\bm{\beta}}$ by
\[
\textrm{Var}(\widehat\mu_{i})=\mbox{tr}\{(\bm{d}_{i}\bm{d}_{i}^{\top})\textrm{Cov}(\widehat{\bm{\beta}})\}.
\]

%
\section{Special models}

Equation~(\ref{bias-beta}) is easily handled algebraically for any type of
nonlinear model, since it involves simple operations on matrices and vectors.
This equation, in conjunction with a computer algebra system such
as {\sf MAPLE} (Abell and Braselton, 1994) will compute
$\bm{B}(\widehat{\bm{\beta}})$ algebraically with minimal effort.
In particular, (\ref{bias-beta}) may simplify considerably if the number of nonlinear
parameters is small. Moreover, for any nonlinear special model, we can calculate the
bias $\bm{B}(\widehat{\bm{\beta}})$ numerically via a software with numerical
linear algebra facilities such as {\sf Ox} (Doornik, 2001) and {\sf R} (R Development
Core Team, 2008).

First, we consider a nonlinear regression model which depends on a single
nonlinear parameter. Equation~(\ref{bias-beta}) gives
\[
B(\widehat{\beta}) = -\frac{2}{\psi_{1}(\alpha)}\frac{\kappa_{2}}{\kappa_{1}^2},
\]
where $\kappa_{1}=\sum_{i=1}^{n}({\rm d}f_{i}/{\rm d}\beta)^2$ and
$\kappa_{2}=\sum_{i=1}^{n}({\rm d}f_{i}/{\rm d}\beta)({\rm d}^2f_{i}/{\rm d}\beta^2)$.
The cons\-tants $\kappa_{1}$ and $\kappa_{2}$ are evaluated at $\widehat{\beta}$
and $\widehat{\alpha}$ to yield $\widehat{B}(\widehat{\beta})$ and the
corrected estimate $\widetilde{\beta}=\widehat{\beta}-\widehat{B}(\widehat{\beta})$.
For example, the simple exponential model $f_{i}=\exp(\beta x_{i})$
leads to $\kappa_{1}=\sum_{i=1}^{n}x_{i}^2\exp(2\beta x_{i})$ and
$\kappa_{2}=\sum_{i=1}^{n}x_{i}^3\exp(2\beta x_{i})$.

As a second example, we consider a partially nonlinear regression model defined by
\begin{equation}\label{Mspecial}
\bm{\mu}= \bm{Z}\bm{\lambda}+\eta\bm{g}(\gamma),
\end{equation}
where $\bm{Z}$ is a known $n\times(p-2)$ matrix of full rank, $\bm{g}(\gamma)$
is an $n\times 1$ vector, $\bm{\beta}=(\bm{\lambda}^{\top},\eta,\gamma)^{\top}$,
$\bm{\lambda}=(\lambda_{1},\ldots,\lambda_{p-2})^{\top}$ and $\eta$ and $\gamma$ are
scalar parameters. This class of models occurs very often in statistical modeling;
see Cook et al.~(1986) and Cordeiro et al.~(2000). For example, $\mu=\lambda_{1}z_{1}+
\lambda_{2}z_{2}+\eta\exp(\gamma x)$ (Gallant, 1975),
$\mu=\lambda -\eta\log(x_{1}+\gamma x_{2})$ (Darby and Ellis, 1976) and
$\mu=\lambda +\eta\log(x_{1}/(\gamma+x_{2}))$ (Stone, 1980). Ratkowsky (1983, Ch.~5)
discusses several models of the form~(\ref{Mspecial}) which include the
asymptotic regression and Weibull-type models given by $\mu=\lambda-\eta\gamma^{x}$ and
$\mu=\lambda-\eta\exp(-\gamma x)$, respectively.

The $n\times p$ local model matrix $\bm{D}$ takes the form $\bm{D}=[\bm{Z},\bm{g}(\gamma),
\eta({\rm d}\bm{g}(\gamma)/{\rm d}\gamma)]$ and, after some algebra, we can obtain
from~(\ref{bias-beta})
\begin{equation}\label{B1}
\bm{B}(\widehat{\bm{\beta}})= -\biggl[\frac{1}{\eta}\Cov(\widehat{\eta}, \widehat{\gamma})\bm{\tau}_{p}
+\frac{\eta}{2}\Var(\widehat{\gamma})\bm{\delta}_{p}\biggr],
\end{equation}
where $\bm{\tau}_{p}$ is a $p\times 1$ vector with a one in the last position and zeros
elsewhere, $\bm{\delta}_{p}=(\bm{D}^{\top}\bm{D})^{-1}\bm{D}^{\top}({\rm d}^2\bm{g}(\gamma)/
{\rm d}\gamma^2$) is simply the set of coefficients from the ordinary regression of the vector
${\rm d}^2\bm{g}(\gamma)/{\rm d}\gamma^2$ on the matrix $\bm{D}$,
and $\Var(\widehat{\gamma})$ and $\Cov(\widehat{\eta},\widehat{\gamma})$ are
the large-sample second moments obtained from the appropriate elements of the
asymptotic covariance matrix
$\Cov(\widehat{\bm{\beta}})=\bm{K}_{\bm{\beta}}^{-1}=(4/\psi_{1}(\alpha))(\bm{D}^{\top}\bm{D})^{-1}$.
It is clear from~(\ref{B1}) that $\bm{B}(\widehat{\bm{\beta}})$ does not depend explicitly
on the linear parameters in $\bm{\lambda}$ and it is proportional to $4/\psi_{1}(\alpha)$.
Further, the covariance term $\Cov(\widehat{\eta},\widehat{\gamma})$ contributes only to the bias of
$\widehat{\gamma}$.

\section{Numerical results}\label{simulation}

We now use Monte Carlo simulation to evaluate the finite-sample performance of the
MLEs of the parameters and of their corrected versions in two nonlinear regression
models. The MLEs of the parameters were obtained by maximizing the log-likelihood
function using the BFGS quasi-Newton method with
ana\-lytical derivatives. This method is generally regarded as the best-performing
nonlinear optimization method (Mittelhammer et al., 2000, p.~199).
The covariate values were selected as random draws from the uniform $\mathcal{U}(0,1)$
distribution and for fixed $n$ those values were kept constant throughout the
experiment. Also, the number of Monte Carlo replications was 10,000.
All simulations were performed using the {\sf Ox}
matrix programming language (Doornik, 2001).\footnote{{\sf Ox} is
freely distributed for academic purposes and available at http://www.doornik.com.}

In order to analyze the performance of the estimates, we computed, for each sample
size and for each estimate: the relative bias (the relative bias of an estimate
$\widehat{\theta}$, defined as $\{\Es(\widehat{\theta})-\theta\}/\theta$,
is obtained by estimating $\Es(\widehat{\theta})$ by Monte Carlo)
and the root mean square error ($\sqrt{{\rm MSE}}$),
where MSE is the estimated mean square error from the 10,000
Monte Carlo replications.

First, we consider the nonlinear regression model
\[
\mu_{i}=\lambda_{1}z_{i1}+\lambda_{2}z_{i2}+\eta\exp(\gamma x_{i}),
\]
where $\varepsilon_{i}\sim\mathcal{SN}(\alpha,0,2)$ for $i=1,\ldots,n$.
The sample sizes were $n= 15, 30$ and 45.
Without loss of generality, the true values of the regression parameters
were taken as $\lambda_{1}=4$, $\lambda_{2}=5$, $\eta=3$, $\gamma=1.5$
and $\alpha=0.5$ and $1.5$.

Table~\ref{tab1} gives the relative biases of both uncorrected and corrected
estimates to show that the bias corrected estimates are much closer to the true
parameters than the unadjusted estimates. For ins\-tance, when $n=15$ and $\alpha=1.5$,
the average of the estimated relative biases for the estimates of the model parameters
is $-0.03244$, whereas the average of the estimated relative biases for
the corrected estimates is $-0.0083$. Hence, the average bias
(in absolute value) of the MLEs is almost four times greater than the average bias
of the corrected estimates. This fact suggests that the second-order bias of
the MLEs should not be ignored in samples of small to moderate size
since they can be non-negligible. The figures in Table 2 show that
the root mean squared errors of the uncorrected and corrected estimates
are very close. Hence, the figures in both tables suggest that the corrected
estimates have good properties.

\begin{table}[!htp]
\caption{Relative biases of the uncorrected and corrected estimates.}\label{tab1}
\begin{tabular}{ccl rrrrr}\hline
$\alpha$  & $n$ & & ${\lambda}_{1}$  & ${\lambda}_{2}$  & ${\eta}$ & ${\gamma}$  & ${\alpha}$ \\\hline
0.5& 15 & MLE & 0.0006 & $-0.0013$ & 0.0011 &  0.0020 & $-0.1691$ \\
  &     & BCE & 0.0007 & $-0.0011$ & 0.0001 &  0.0008 & $-0.0395$ \\\cline{2-8}
  & 30  & MLE & 0.0001 & $-0.0013$ & 0.0013 &  0.0009 & $-0.0811$ \\
  &     & BCE & 0.0002 & $-0.0012$ & 0.0007 &$-0.0001$& $-0.0092$ \\\cline{2-8}
  & 45  & MLE & 0.0003 & $-0.0012$ & 0.0007 &  0.0008 & $-0.0537$ \\
  &     & BCE & 0.0003 & $-0.0011$ & 0.0003 &  0.0001 & $-0.0042$ \\\hline

1.5& 15 & MLE & $-0.0068$ & $-0.0083$ & 0.0248 &  0.0197 & $-0.1916$ \\
  &     & BCE & $-0.0055$ & $-0.0046$ & 0.0113 &  0.0056 & $-0.0481$ \\\cline{2-8}
  & 30  & MLE & $-0.0016$ & $-0.0034$ & 0.0079 &  0.0078 & $-0.0933$ \\
  &     & BCE & $-0.0011$ & $-0.0018$ & 0.0027 &  0.0012 & $-0.0116$ \\\cline{2-8}
  & 45  & MLE & $-0.0028$ & $-0.0027$ & 0.0052 &  0.0026 & $-0.0614$ \\
  &     & BCE & $-0.0023$ & $-0.0018$ & 0.0023 &$-0.0005$& $-0.0048$ \\\hline
\multicolumn{8}{l}{{\small BCE: bias corrected estimate.}}
\end{tabular}
\end{table}

\begin{table}[!htp]
\caption{Root mean squared errors of the uncorrected and corrected estimates.}\label{tab2}
\begin{tabular}{ccl rrrrr}\hline
$\alpha$  & $n$ & & ${\lambda}_{1}$  & ${\lambda}_{2}$  & ${\eta}$ & ${\gamma}$  & ${\alpha}$ \\\hline
0.5& 15 & MLE & 0.4093 & 0.4920 & 0.2707 & 0.0924  & 0.1234 \\
  &     & BCE & 0.4093 & 0.4921 & 0.2709 & 0.0922  & 0.1067 \\\cline{2-8}
  & 30  & MLE & 0.3006 & 0.3806 & 0.2113 & 0.0688  & 0.0763 \\
  &     & BCE & 0.3006 & 0.3806 & 0.2114 & 0.0686  & 0.0702 \\\cline{2-8}
  & 45  & MLE & 0.2434 & 0.2874 & 0.1768 & 0.0567  & 0.0590 \\
  &     & BCE & 0.2434 & 0.2874 & 0.1769 & 0.0566  & 0.0555 \\\hline

1.5& 15 & MLE & 1.6302 & 1.1230 & 0.9756 & 0.3235  & 0.3938 \\
  &     & BCE & 1.6333 & 1.1274 & 0.9819 & 0.3152  & 0.3315 \\\cline{2-8}
  & 30  & MLE & 0.9684 & 0.7003 & 0.5785 & 0.1931  & 0.2399 \\
  &     & BCE & 0.9693 & 0.7011 & 0.5807 & 0.1908  & 0.2155 \\\cline{2-8}
  & 45  & MLE & 0.6505 & 0.5575 & 0.3895 & 0.1318  & 0.1837 \\
  &     & BCE & 0.6507 & 0.5577 & 0.3901 & 0.1311  & 0.1700 \\\hline
\multicolumn{8}{l}{{\small BCE: bias corrected estimate.}}
\end{tabular}
\end{table}

When the parameter $\alpha$ increases, the finite-sample performance of
the MLEs deteriorates (see Tables~\ref{tab1} and~\ref{tab2}). For instance,
when $n=15$, the relative biases of $\widehat{\gamma}$ (MLE) and $\widetilde{\gamma}$ (BCE)
were 0.0020 and 0.0008 (for $\alpha=0.5$) and 0.0197 and 0.0056 (for $\alpha=1.5$),
which indicate an increase in the relative biases of nearly 10 and 7 times,
respectively. Also, the root mean squared errors in the same order
were 0.0924 and 0.0922 (for $\alpha=0.5$) and 0.3235 and 0.3152 (for
$\alpha=1.5$).

Next, we consider the very known Michaelis--Menton model, which is very
useful for estimating growth curves, where it is common for the response
to approach an asymptote as the stimulus increases. The Michaelis--Menton
model (Ratkowsky, 1983) provides an hyperbolic form for $\mu_{i}$ against $x_{i}$ given by
\[
\mu_{i} = \frac{\eta x_{i}}{\gamma + x_{i}},\quad i = 1,2,\ldots,n,
\]
where the curve has an asymptote at $\mu = \eta$.
Here, the sample sizes were $n= 20, 30, 40$ and 50. Also,
the true values of the regression parameters
were taken as $\eta = 3$ and $\gamma= 0.5$, with $\alpha=0.5$.

Table~\ref{tab3} gives the relative biases and root mean squared errors
of the uncorrected and corrected estimates. The figures in this table
reveal that the MLEs of the parameters can be substantially biased,
even when $n = 50$, and that the bias correction is very effective.
In terms of MSE, the adjusted estimates are slightly better than the
ordinary MLEs.

\begin{table}[!htp]
\caption{Relative biases and root mean squared errors of the uncorrected and corrected estimates; $\alpha = 0.5$
and different sample sizes.}\label{tab3}
\begin{tabular}{cl rrr|rrr}\hline
& & \multicolumn{3}{c|}{Relative Bias}  & \multicolumn{3}{c}{$\sqrt{{\rm MSE}}$} \\\cline{3-8}
$n$ & & $\eta$  & $\gamma$  & $\alpha$ & $\eta$  & $\gamma$  & $\alpha$ \\\hline
 20 & MLE &  0.0476  &   0.1718  & $-0.0669$ & 0.6984 & 0.3947 & 0.0859 \\
    & BCE &$-0.0016$ & $-0.0081$ & $-0.0061$ & 0.5264 & 0.2783 & 0.0847 \\\hline

 30 & MLE &  0.0313  &  0.1077  & $-0.0439$ & 0.5245 & 0.2750 & 0.0684 \\
    & BCE & $0.0004$ & $0.0012$ & $-0.0024$ & 0.4478 & 0.2252 & 0.0678 \\\hline

 40 & MLE &  0.0215  &   0.0754  & $-0.0330$ & 0.4222 & 0.2207 & 0.0582 \\
    & BCE &$-0.0001$ & $-0.0003$ & $-0.0015$ & 0.3835 & 0.1954 & 0.0578 \\\hline

 50 & MLE & 0.0160  &   0.0558  & $-0.0259$ & 0.3609 & 0.1862 & 0.0516 \\
    & BCE &$0.0000$ & $-0.0001$ & $-0.0005$ & 0.3380 & 0.1710 &  0.0514 \\\hline

\multicolumn{8}{l}{{\small BCE: bias corrected estimate.}}
\end{tabular}
\end{table}

\section{Application}\label{application}

Obviously, due to the genesis of the Birnbaum--Saunders distribution, the
fatigue processes are by excellence ideally modeled by this model. We
now consider an application to a biaxial fatigue data set
reported by Rieck and Nedelman (1991) on the life of a metal piece in
cycles to failure. The response $N$ is the number of cycles to failure and
the explanatory variable $w$ is the work per cycle (mJ/m$^3$). The data of
forty six observations were taken from Table 1 of Galea et al.~(2004).
\begin{figure}
\centering
\includegraphics[width=12cm, height=9cm]{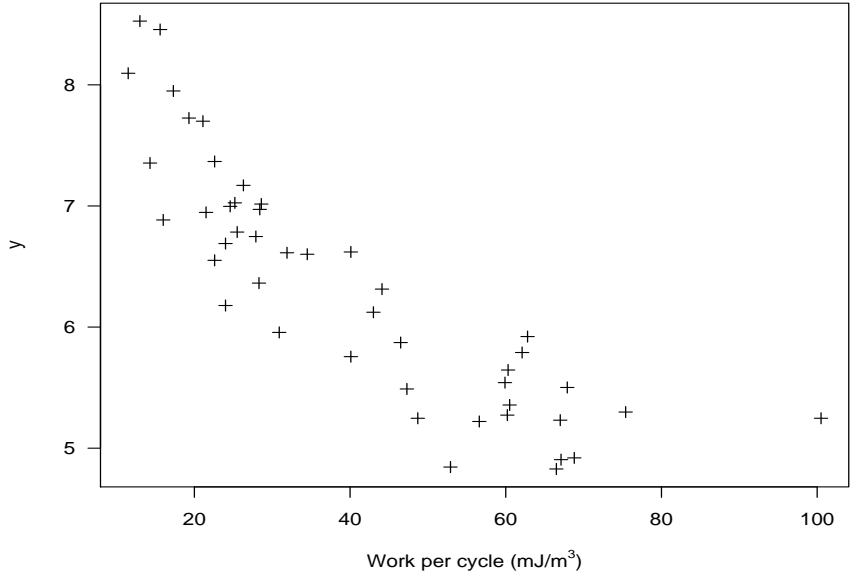}
\caption{Scatter-plot of the data set.}\label{fig1}
\end{figure}

Rieck and Nedelman (1991) proposed the following model for the
biaxial fatigue data:
\begin{equation}\label{RNmodel}
y_{i}=\beta_{1}+\beta_{2}\log w_{i}+ \varepsilon_{i},
\end{equation}
where $y_{i} = \log N_{i}$ and $\varepsilon_{i}\sim\mathcal{SN}(\alpha, 0, 2)$,
for $i = 1,\ldots,46$. The MLEs (the corresponding standard errors in parentheses)
are: $\widehat{\beta}_{1} = 12.2797$\,(0.3942),
$\widehat{\beta}_{2} = -1.6708$\,(0.1096) and $\widehat{\alpha}=0.4104$\,(0.0428).
We take the logarithm of $w$ to ensure a linear relationship between
the response variable ($y$) and the covariate in~(\ref{RNmodel});
see Galea et al.~(2004, Figure 1). However, Figure~\ref{fig1} suggests a
nonlinear relationship between the response variable and the covariate $w$.

Here, we proposed the nonlinear regression model
\begin{equation}\label{Nonlinear}
y_{i}=\beta_{1}+\beta_{2}\exp(\beta_{3}/w_{i})+\varepsilon_{i},\quad i=1,\ldots,46,
\end{equation}
where $\varepsilon_{i}\sim\mathcal{SN}(\alpha,0,2)$. The MLEs
(the standard errors in parentheses) are:
$\widehat{\beta}_{1} = 8.9876$\,(0.7454), $\widehat{\beta}_{2}= -5.1802$\,(0.5075),
$\widehat{\beta}_{3} = -22.5196$\,(7.3778) and $\widehat{\alpha} = 0.40$\,(0.0417).
The bias corrected estimates are:
$\widetilde{\beta}_{1} = 8.7806$\,(0.7734), $\widetilde{\beta}_{2} = -4.9362$\,(0.5266),
$\widetilde{\beta}_{3} = -22.1713$\,(7.6548) and $\widetilde{\alpha} = 0.4157$\,(0.0433).
Hence, the uncorrected estimates are slightly different from the bias corrected
estimates even for large samples ($n=46$ observations).

\begin{figure}
\centering
\includegraphics[width=12cm, height=9cm]{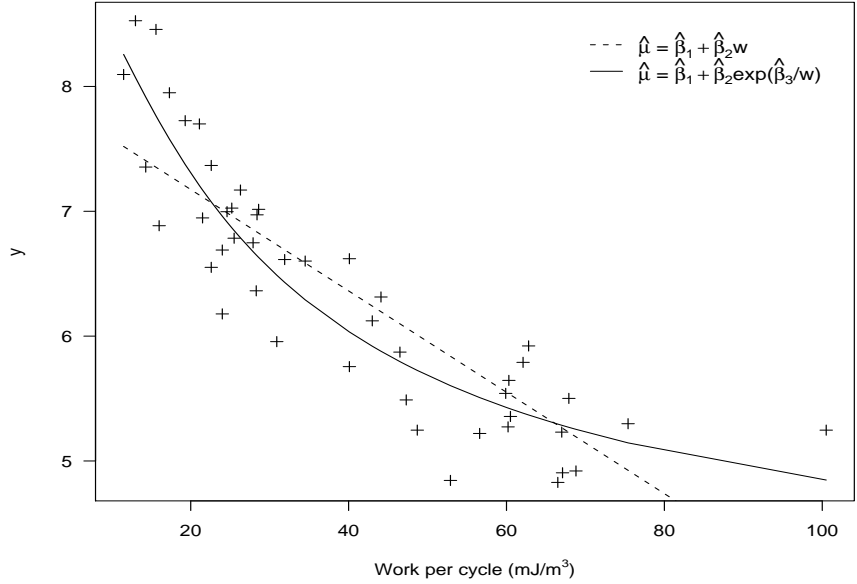}
\caption{Scatter-plot and the fitted models.}\label{fig2}
\end{figure}

Figure~\ref{fig2} gives the scatter-plot of the data, the
fitted model~({\ref{Nonlinear}}) and the fitted straight line,
say $y_{i}=\beta_{1}+\beta_{2}w_{i}+\varepsilon_{i}$, where the MLEs
are: $\widehat{\beta}_{1}=7.9864$\,(0.1622), $\widehat{\beta}_{2}=-0.0406$\,(0.0036)
and $\widehat{\alpha}= 0.52$\,(0.0542). Figure~\ref{fig2} shows that the nonlinear
model (10) (unlike the linear model) fits satisfactorily to the
fatigue data. The $46$th observation (the one with work per cycle near 100)
can be an influential data. However, it is not possible to say whether this
observation is influential or not without using an efficient way to detect
influential observations in the new class of models. Influence diagnostic analysis
for this class of models will be developed in future research.

\begin{figure}
\centering
\includegraphics[width=12cm, height=9cm]{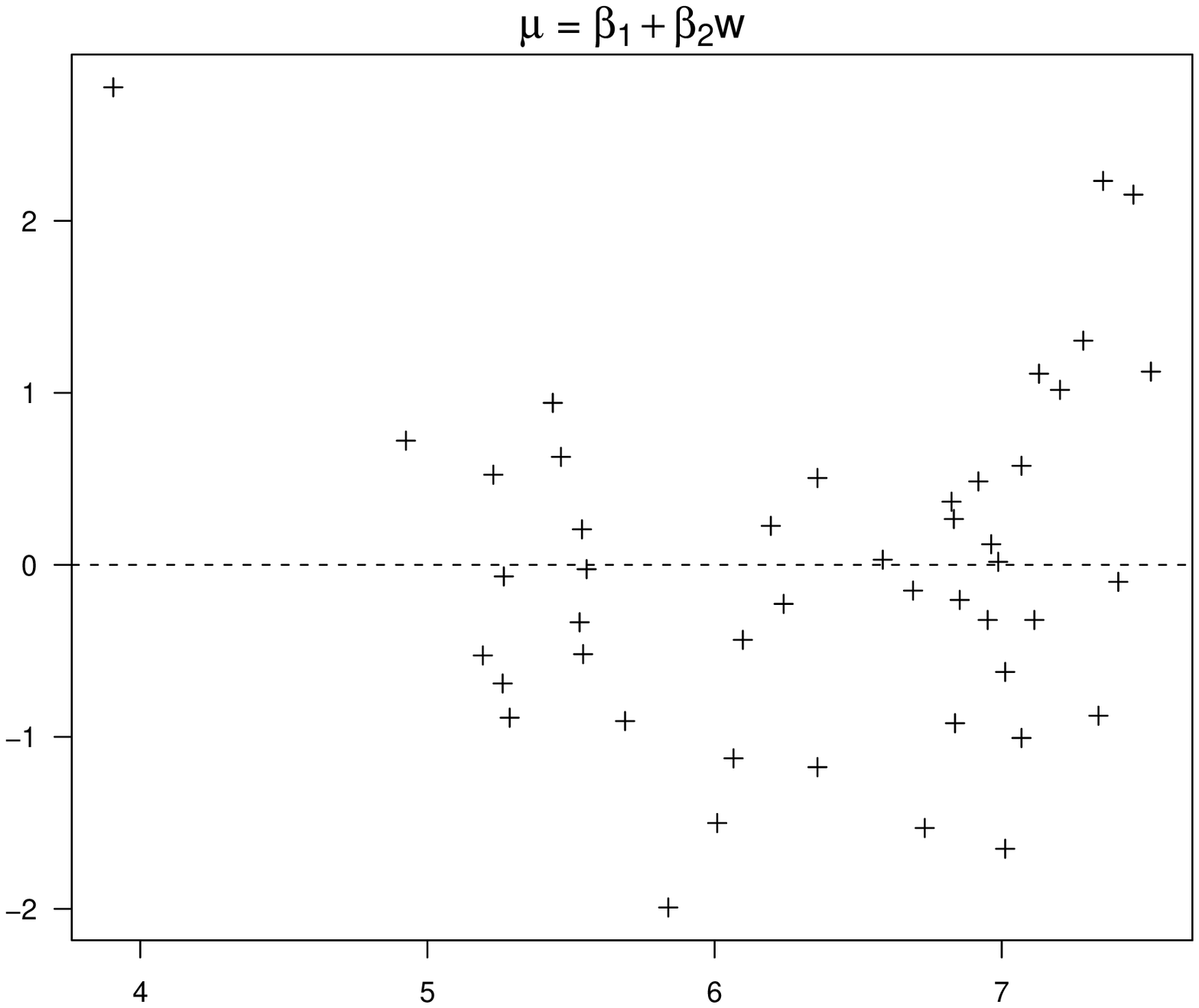}
\includegraphics[width=12cm, height=9cm]{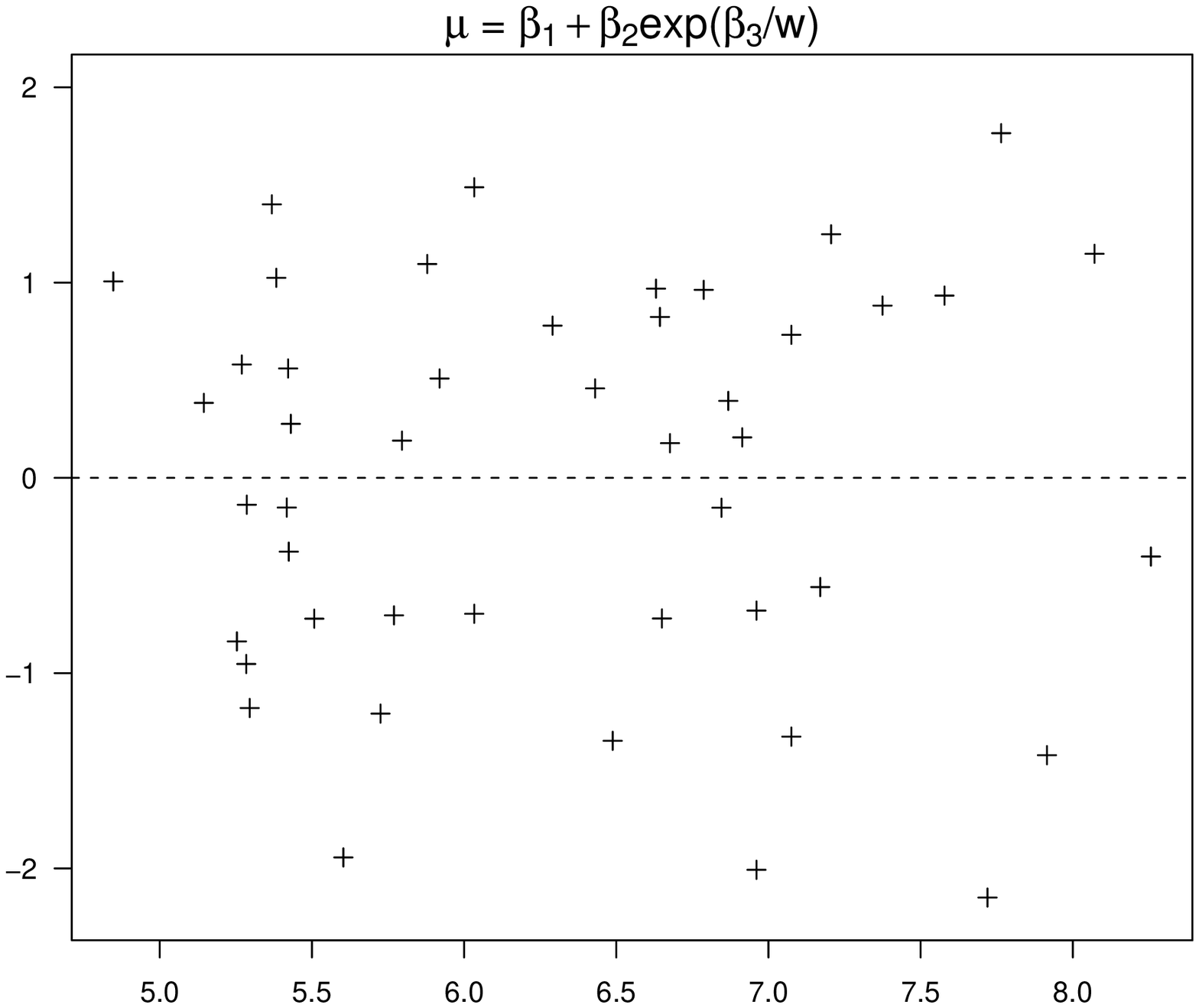}
\caption{Index plot of $\widehat{R}_{i}$ versus $\widehat{\mu}_{i}$.}\label{fig3}
\end{figure}

Following Xie and Wei (2007), we obtain the residuals
$\widehat{\varepsilon}_{i}=y_{i}-\widehat{\mu}_{i}$ and $\widehat{R}_{i} = 2
\widehat{\alpha}^{-1}\sinh(\widehat{\varepsilon}_{i}/2)$. Figure~\ref{fig3}
gives the scatter-plot of $\widehat{R}_{i}$ versus the predicted values
$\widehat{\mu}_{i}$ for both fitted models:
(i) $y_{i}=\beta_{1} + \beta_{2}w_{i} + \varepsilon_{i}$; and
(ii) $y_{i}=\beta_{1} + \beta_{2}\exp(\beta_{3}/w_{i}) + \varepsilon_{i}$.
Figure~\ref{fig3} shows that the distribution of $\widehat{R}_{i}$ is
approximately normal for model (ii) but this is not true for model (i).
Based upon the fact that $U\sim\mathcal{SN}(\alpha,\mu,\sigma)$ if
$2\alpha^{-1}\sinh\{(U-\mu)/\sigma\}\sim\mathcal{N}(0,1)$, then the
residual $\widehat{\varepsilon}_{i}$ should follow approximately
a sinh-normal distribution.

%
\section{Conclusions}\label{conclusions}

The Birnbaum--Saunders distribution is widely used to model times to fai\-lure for
materials subject to fatigue. The purpose of the paper was two fold. First, we propose
a new class of Birnbaum--Saunders nonlinear regression mo\-dels which generalizes
the regression model described in Rieck and Nedelman (1991). Second, we give
simple formulae for calculating bias corrected ma\-ximum likelihood estimates
of the parameters of these models. The simulation results
presented show that the bias correction derived is very effective, even when the
sample size is large. Indeed, the bias correction mechanism adopted yields
adjusted maximum likelihood estimates which are nearly unbiased. We also present
an application to a real fatigue data set that illustrates the usefulness of the
proposed model. Future research will be devoted to a study of diagnostics and
influence analysis in the new class of nonlinear models.

\section*{Acknowledgments}

We gratefully acknowledge grants from FAPESP and CNPq (Brazil). The authors are also
grateful to an associate editor and two referees for helpful comments and
suggestions.

%

\end{document}